\documentclass
[twocolumn,prb,showpacs,preprintnumbers,amsmath,amssymb]{revtex4-2}%
\usepackage{graphicx}
\usepackage{graphics}
\usepackage{dcolumn}
\usepackage{bm}
\usepackage{amsmath}
\usepackage{amssymb}
\usepackage{wasysym}
\usepackage{mathrsfs}
\usepackage{gensymb}
\usepackage{subfigure}
\usepackage{color,soul}
\usepackage{MnSymbol}
\usepackage{hyperref}
\usepackage{amsfonts}%
\providecommand{\U}[1]{\protect\rule{.1in}{.1in}}
\definecolor{blue}{rgb}{0,0,1}
\definecolor{red}{rgb}{1,0,0}
\definecolor{darkred}{rgb}{0.5,0,0}
\definecolor{greay}{rgb}{0.432,0.431,0.296}
\begin{document}

\title{Magnetic order of Dy$^{3+}$ and Fe$^{3+}$ moments in antiferromagnetic DyFeO$_{3}$ probed by
spin Hall magnetoresistance and spin Seebeck effect}
\author{G. R. Hoogeboom$^{1}$\email{Electronic mail: g.r.hoogeboom@gmail.nl}, T. Kuschel$^{2}$, G.E.W. Bauer$^{1,3}$, M. V.
Mostovoy$^{1}$, A. V. Kimel$^{4}$ and B. J. van Wees$^{1}$}
\affiliation{$^1$ Physics of Nanodevices, Zernike Institute for Advanced Materials, University of Groningen, Nijenborgh 4, 9747 AG Groningen, The Netherlands.\\
	$^2$Center for Spinelectronic Materials and Devices, Department of Physics, Bielefeld University, Universit\"atsstra$\beta$e 25, 33615 Bielefeld, Germany.\\
		$^3$AIMR \& Institute for Materials Research, Tohoku University, Aoba-ku, Katahira 2-1-1, Sendai, Japan \\
	$^4$Spectroscopy of Solids and Interfaces, Institute of Molecules and Materials, Radboud University Nijmegen, Heyendaalseweg 135, 6525 AJ
	Nijmegen, The Netherlands} \ 
\date{\today}

\begin{abstract}
We report on spin Hall magnetoresistance (SMR) and spin Seebeck effect (SSE) in
single crystal of the rare-earth antiferromagnet DyFeO$_{3}$ with a thin Pt
film contact. The angular shape and symmetry of the SMR at elevated temperatures reflect the
antiferromagnetic order of the Fe$^{3+}$ moments as governed by the Zeeman
energy, the magnetocrystalline anisotropy and the Dzyaloshinskii-Moriya
interaction. We interpret the observed linear dependence of the signal on the
magnetic field strength as evidence for field-induced order of the Dy$^{3+}$ moments up to room temperature. At and below the Morin temperature of 50$\,$K, the SMR monitors the spin-reorientation phase transition of Fe$^{3+}$ spins.
Below 23$\,$K,  additional features emerge that persist below 4$\,$K, the
ordering temperature of the Dy$^{3+}$ magnetic sublattice.
We conclude that the combination of SMR and SSE is a simple and efficient tool
to study spin reorientation phase transitions and sublattice
magnetizations.\newline

\end{abstract}
\maketitle


\section{Introduction}

\label{Introduction}

Antiferromagnets (AFMs) form an abundant class of materials that offer many
advantages over ferromagnets (FMs) for applications in high-density magnetic
logics and data storage devices. AFMs support high-frequency dynamics in the
THz regime that allows faster writing of magnetic bits compared to
FMs. The absence of magnetic stray fields minimizes on-chip
cross-talk and allows downsizing devices that are robust against magnetic
perturbations \cite{Loth2012}. On the other hand, most magnetic detection
methods observe only the FM order. Recent developments in the detection \cite{Hoogeboom2017} and manipulation \cite{Afanasiev2016,Wadley2016,Moriyama2018} of the AFM order reveal its many opportunities. 

The AFM DyFeO$_{3}$ (DFO) belongs to a family of rare-earth
transition metal oxides called orthoferrites that display many unusual
phenomena such as weak ferromagnetism (WFM), spin-reorientation transitions,
strong magnetostriction, multiferroicity including a large linear
magnetoelectric effect \cite{Tokunaga2008}. Their magnetic properties are
governed by the spin and orbital momenta of 4f rare-earth ions coupled to
the magnetic moment of 3d transition metal ions.

The magnetization of dielectrics can be detected electrically by the spin Hall
magnetoresistance (SMR) in heavy metal contacts with a large spin Hall angle
such as Pt \cite{Nakayama2013}. This phenomenon is sensitive to FM, but also
AFM spin order \cite{Hoogeboom2017, Fischer2018, Jiang2018, Lebrun2019}. With
a Pt contact, information about AFMs can be also retrieved by the spin Seebeck
effect (SSE) under a temperature gradient \cite{Wu2016, Rezende2017,
Hoogeboom2020}.

Here, we track the field-dependence of the coupled Dy$^{3+}$ and Fe$^{3+}$
magnetic order\textit{ }
as a function of temperature by both SMR and SSE. A sufficiently strong
magnetic field in the $ab$ plane of DFO forces the N\'{e}el vector to follow a
complex path out of the $ab$ plane. A theoretical spin model explains the
observations in terms of Fe$^{3+}$ spin rotations that are governed by the
competition between the magnetic anisotropy, Zeeman energy, and Dzyaloshinskii-Moriya interaction (DMI). 
The Dy$^{3+}$ moments are disordered at room temperature but
nevertheless affect the magnitude of the SMR. At the so-called Morin phase
transition at $\sim50\,\mathrm{K}$ the Fe$^{3+}$ spins rotate by $90{^{\circ}}$, 
causing a step-like anomaly in the SMR. At even lower temperatures, we
observe two separate features tentatively assigned to the re-orientation of
Fe$^{3+}$ spins in an applied magnetic field and another related to the
ordering of Dy$^{3+}$ orbital moments. Below 23$\,$K, the SMR signal is $\sim$
1\%, 1-2 orders of magnitude larger than reported for other materials
\cite{Nakayama2013,Hoogeboom2017}. Both Fe$^{3+}$ and Dy$^{3+}$ moments appear
to contribute to the SSE; a magnetic field orders the Dy$^{3+}$ moments and
suppresses the Fe$^{3+}$ contribution. The complex SMR and SSE is evidence of
a coupling between the Fe$^{3+}$ and Dy$^{3+}$ magnetic subsystems.

The paper is organized as follows. In Section \ref{Properties} we review the
magnetic and multiferroic properties of DFO. The theory of the magnetic
probing methods are discussed in Sec. \ref{Theory} with Subsec. \ref{SMR} the SMR and Subsec. \ref{Spin Seebeck effect} the SSE. In Subsec. \ref{Fabrication}, the fabrication, characterization and measurement
techniques are explained. Further, a model including the DMI, Zeeman energy
and magnetic anisotropy is employed in Subsec. \ref{Model}. The SMR results at elevated temperatures including the model fits as well as SMR and SSE results at low temperatures are described and discussed in Sec. \ref{Results}.


\section{Magnetic and multiferroic properties of DFO}

\label{Properties}

\begin{figure}[t]
\includegraphics[width=8.4cm]{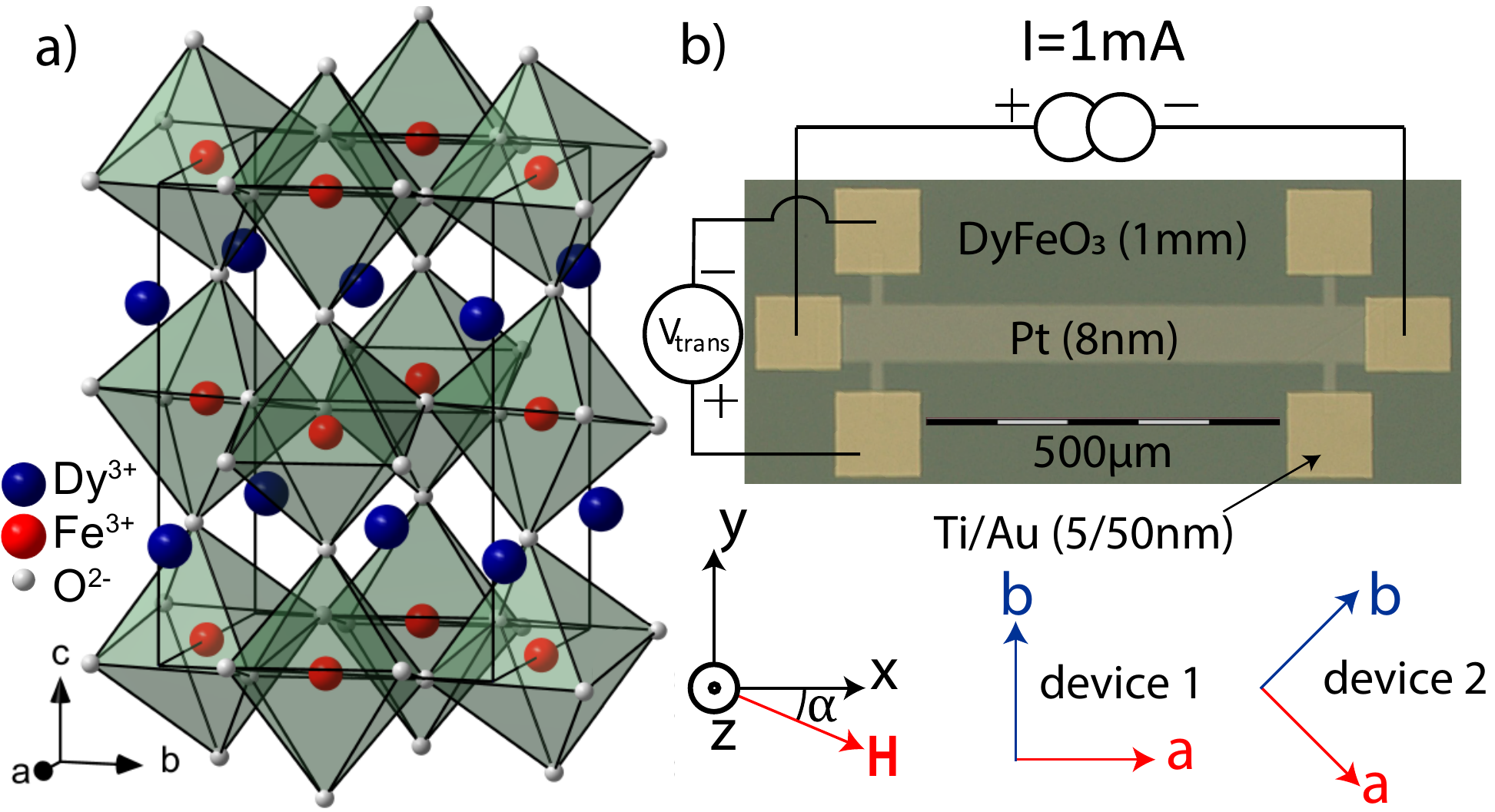} \caption{(a) DFO crystal 
unit cell. The blue, red and white spheres represent Dy$^{3+}$,
Fe$^{3+}$ and O$^{2-}$ ions, respectively. (b)~Optical image of the Pt Hall bar
on top of the bulk DFO crystal. The lines indicate voltage
probes, AC source and angle $\alpha$ of the external magnetic field $\mathbf{H}$. In the
two devices, the crystallographic directions $a$ and $b$ are rotated by
$45{^{\circ}}$ in the $xy$ plane, the reference frame of the Hall bar. }%
\label{fig:sample}%
\end{figure}

DFO is a perovskite with an orthorhombic ($D_{2h}^{16}$ - Pbnm)
crystallographic structure. It consists of alternating Fe$^{3+}$ and Dy$^{3+}$
$ab$ planes, in which the Fe ions are located inside O$^{2-}$ octahedrons
(Fig. \ref{fig:sample}a)). The large Dy$^{3+}$ magnetic moments ($J=15/2$)
order at a low temperature, $T_{\mathrm{N}}^{\mathrm{Dy}}=4\,$K. The high
N\'{e}el temperature $T_{\mathrm{N}}^{\mathrm{Fe}}=645~$K indicates strong
inter- and intra-plane AFM Heisenberg superexchange between the Fe$^{3+}$
magnetic moments ($S=5/2$). The AFM order of the Fe moments is of the G-type
with N\'{e}el vector $\mathbf{G}$ (anti)parallel to the crystallographic $a$
axis ($\Gamma_{4}$ symmetry \cite{turov1963}). The broken inversion symmetry
enables a DMI \cite{DZYALOSHINSKY1958,Moriya1960} that in the $\Gamma
_{4}$-phase causes a WFM $\mathbf{m}_{\mathrm{WFM}}\Vert\mathbf{c}$ by the small ($\sim0.5{^{\circ}}${)} canting of the Fe spins \cite{turov1963}.

A first-order Morin transition from the WFM $\Gamma_{4}$-phase to the purely
AFM $\Gamma_{1}$-phase occurs when lowering the temperature below 50$\,$K. At
this transition, the direction of the magnetic easy axis abruptly changes from
the $a$- to the $b$-direction. A magnetic field higher than a critical
magnetic field, $H_{\mathrm{cr}}$, along the $c$ axis re-orients the N\'{e}el
vector back to the $a$ axis and recovers the $\Gamma_{4}$-phase. Below
$T_{\mathrm{N}}^{\mathrm{Dy}}$, the Dy$^{3+}$ moments form a noncollinear
Ising-like AFM order with Ising axes rotated by $\pm33{^{\circ}}$ from the $b$
axis \cite{Holmes1972} that corresponds to a $G_{a}^{\prime}A_{b}^{\prime}$
state in Bertaut's notation \cite{Bertaut1963}. The simultaneous presence of
ordered Fe and Dy magnetic moments breaks inversion symmetry and, under an
applied magnetic field, induces an electric polarization \cite{Yamaguchi1973}
by exchange striction that couples the Fe and Dy magnetic sublattices
\cite{Tokunaga2008, Stroppa2010}. Higher magnetic fields destroy the AFM order
of the Dy$^{3+}$ moments and thereby the electric polarization \cite{Wang2016}%
.\newline

Spins in this material can be controlled by light through the inverse Faraday
effect \cite{Afanasiev2016}, as well as by temperature and magnetic field.
Re-orientation of the Fe moments has been studied by magnetometry
\cite{Belov1974}, Faraday rotation \cite{Szymczak1977}, M\"{o}ssbauer
spectroscopy \cite{Prelorendjo1980} and neutron scattering measurements
\cite{Wang2016}. The Morin transition at 50$\,$K causes large changes in the
specific heat \cite{Zhang2015} and entropy \cite{Ke2015}.

\section{probing methods}
\label{Theory}

\subsection{Spin Hall magnetoresistance}
\label{SMR}

The SMR is caused by the spin-charge conversion
in a thin heavy metal layer in contact with a magnet \cite{Bauer2013}. The
spin Hall effect induces a spin current transverse to an applied charge
current and thereby an electron spin accumulation at surfaces and interfaces.
Upon reflection at the interface to a magnetic insulator, electrons experience
an exchange interaction that depends on the angle between their spin
polarization and that of the interface magnetic moments, while the latter can
be controlled by an applied magnetic field. The reflected spin current is
transformed back into an observable charge current by the inverse spin Hall
effect. The interface exchange interaction is parameterized by the complex
spin mixing conductance. The result is a modulation of the charge transport
that depends on the orientation of the applied current and the interface
magnetic order. In a Hall bar geometry, this affects the longitudinal
resistance and causes a planar Hall effect, i.e. a Hall voltage even when the
magnetic field lies in the transport plane.

SMR is a powerful tool to investigate the magnetic ordering at the interface
of collinear \cite{Nakayama2013, Bauer2013, Vlietstra2013, Althammer2013} and
noncollinear ferrimagnets \cite{Ganzhorn2016, Dong2018} as well as spin
spirals \cite{Aqeel2015, Aqeel2016}. Recently, a \textquotedblleft
negative\textquotedblright\ SMR has been discovered for AFMs
\cite{Hoogeboom2017, Fischer2018, Jiang2018, Lebrun2019}, i.e. an SMR with a
$90{^{\circ}}$ phase shift of the angular dependence as compared to
FMs, which shows that the AFM N\'{e}el vector
$\mathbf{G}$ tends to align itself normal to the applied magnetic field. The observable in
AFMs is therefore the N\'{e}el vector rather than the net magnetization
\cite{Hoogeboom2017}.

The longitudinal and transverse electrical resistivities $\rho_{L}$ and
$\rho_{T}$ of Pt on an AFM read \cite{Hoogeboom2017}
\begin{equation}
\rho_{L}=\rho+\Delta\rho_{0}+\Delta\rho_{1}(1-G_{y}^{2}) \label{rhol}%
\end{equation}%
\begin{equation}
\rho_{T}=\Delta\rho_{1}G_{x}G_{y}+\Delta\rho_{2}m_{z}+\Delta\rho
_{\mathrm{Hall}}H_{z} \label{rhot}%
\end{equation}
with $G_{i}$ and $H_{i}$ with $i\in\left\{  x,y,z\right\} $ as the
Cartesian components of the (unit) N\'{e}el and the applied magnetic field
vectors, respectively. $m_{z}$ is the out-of-plane (OOP) component of the unit
vector in the direction of the WFM magnetization. $\Delta\rho_{0}$ is an
angle-independent interface correction to the bulk resistivity $\rho$.
$\Delta\rho_{\mathrm{Hall}}H_{z}$ is the ordinary Hall resistivity of Pt in
the presence of an OOP component of the magnetic field.
$\Delta\rho_{1}\left(  \Delta\rho_{2}\right)$ is proportional to the real
(imaginary) part of the interface spin-mixing conductance. $\Delta\rho_{2}$ is a resistance induced by the effective WFM field, believed to be small in most circumstances.

The interface Dy$^{3+}$ moments can contribute to the SMR when ordered. 
Below $T_{\mathrm{N}}^{\mathrm{Dy}}$, the Dy$^{3+}$ moments are AFM aligned with
N\'{e}el vector $\mathbf{G}^{\mathrm{Dy}}$. Above $T_{\mathrm{N}}%
^{\mathrm{Dy}}$ and in sufficiently large applied magnetic fields, the
Dy$^{3+}$ moments contribute to the SMR in Eqs. (\ref{rhol},\ref{rhot})
after replacing the N\'{e}el vector $\mathbf{G}^{\mathrm{Dy}}$ by the (nearly
perpendicular) magnetization $\mathbf{m}^{\mathrm{Dy}}$.
Disregarding magnetic anisotropy and DMI for the moment, the spin mixing conductance term $\Delta\rho
_{1}\,m_{x}^{\mathrm{Dy}}m_{y}^{\mathrm{Dy}}$ phase-shifts the SMR by $90{^{\circ}}$ relative to the pure 
 AFM contribution. The term $\Delta\rho_{2}m_{z}$ changes sign with $m_{z}$ and its contribution $\sim H_{z}$ cannot be distinguished from the ordinary Hall effect $\Delta\rho_{\mathrm{Hall}}H_{z}$ in Pt. We remove a linear magnetic field dependence from the OOP SMR measurements. Residual non-linear effects from $\Delta\rho_{2}m_{z}$ may persist, but should be small in the $\Gamma_{4}$ phase. A finite $\Delta\rho_{2}m_{z}$ has been
reported in conducting AFMs \cite{Zhao2019}, but we do not observe a
significant contribution down to 60$\,$K. \newline

\subsection{Spin Seebeck effect}

\label{Spin Seebeck effect}

A heat current in a FM excites a spin current that in insulators is
carried mainly by magnons, the quanta of the spin wave excitations of the
magnetic order. We can generate a temperature bias simply by the Joule heating
of a charge current in a metal contact. A magnon flow $\mathbf{j}_{m}$ can
also be generated by a gradient of a magnon accumulation or chemical potential
$\mu_{m}$ \cite{Cornelissen2016}. Therefore
\begin{equation}
\mathbf{j}_{m}=-\sigma_{m}(\mathbf{\nabla}\mu_{m}+S_{S}\mathbf{\nabla}T)
\label{SSE}%
\end{equation}
with $\sigma_{m}$ as the magnon spin conductivity and $S_{S}$ the spin
Seebeck coefficient. Thermal magnons can typically diffuse over several
$\mathrm{\mu m}$ \cite{Cornelissen2015,Lebrun2018a,Oyanagi2019}, which implies
that the SSE mainly probes bulk rather than interface magnetic properties. The
magnons in simple AFMs typically come in degenerate pairs with opposite
polarization that split under an applied magnetic field \cite{Cheng2014,
Wu2016}. The associated imbalance of the magnon populations cause a non-zero
spin Seebeck effect \cite{Hoogeboom2020}. Paramagnets display a field-induced
SSE effect \cite{Oyanagi2019} for the same reason, so aligned Dy$^{3+}$
moments can contribute to an SSE in DFO. A magnon accumulation at the
interface to Pt injects a spin current $\mathbf{j}_{s}$ that can be observed
as an inverse spin Hall effect voltage $\mathbf{V}_{ISHE}=\rho\theta
_{SH}(\mathbf{j}_{s}\times\boldsymbol{\sigma})$, where $\theta_{SH}$ is the
spin Hall angle and $\boldsymbol{\sigma}$ is the spin polarization. The SMR
and SSE can be measured simultaneously by a lock-in technique
\cite{Vlietstra2014}.

\section{Methods}

\label{Methods}

\subsection{Fabrication, characterization and measurements}

\label{Fabrication}

We confirmed the crystallographic direction of our single crystal by X-ray
diffraction before sawing it into slices along the \textit{ab} plane and
polishing them. Two devices were fabricated on different slices of the
materials using a three step electron beam lithography process; markers were
created to align the devices along two different crystallographic directions.
After fabrication of an 8$\,$nm thick Pt Hall bar, 50$\,$nm Ti/Au contact pads
were deposited.

The angular dependence of the magnetoresistance below 50$\,$K is complex and
hysteretic. Phase changes are associated by internal strains that can cause
cracks in the bulk crystal. We therefore carried out magnetic field sweeps at low
temperatures very slowly, with a waiting time of 60 seconds between each field
step. The response was measured with a 1$\,$mA (100$\,\mathrm{\mu}$A) AC
current through the Pt Hall bar in device 1 (device 2) with a frequency of
7.777$\,$Hz. The first and second harmonic transverse and longitudinal lock-in
voltages as measured with a superconducting magnet in a cryostat with variable
temperature insert are the SMR and SSE effects, respectively.

Below the transition temperature, the Morin transition is induced by a
magnetic field along the $c$ axis that rotates the N\'{e}el vector from
$\mathbf{a}$ to $\mathbf{b}$. For device 1, this does not change the transverse
resistance since $G_{x}^{\mathrm{Fe}}G_{y}^{\mathrm{Fe}}=0$ when the N\'{e}el
vector is in either the $x$- or $y$-direction. On the other hand, device 2 is
optimized for the observation of the Morin transition, because, as discussed
below, the transverse resistance should be maximally positive when
$\mathbf{G}\Vert\mathbf{b}$ and maximally negative when $\mathbf{G}%
\Vert\mathbf{a}$. \newline

\subsection{Modelling the SMR of Pt$|$DFO}

\label{Model}

\begin{figure}[t]
\includegraphics[width=8.4cm]{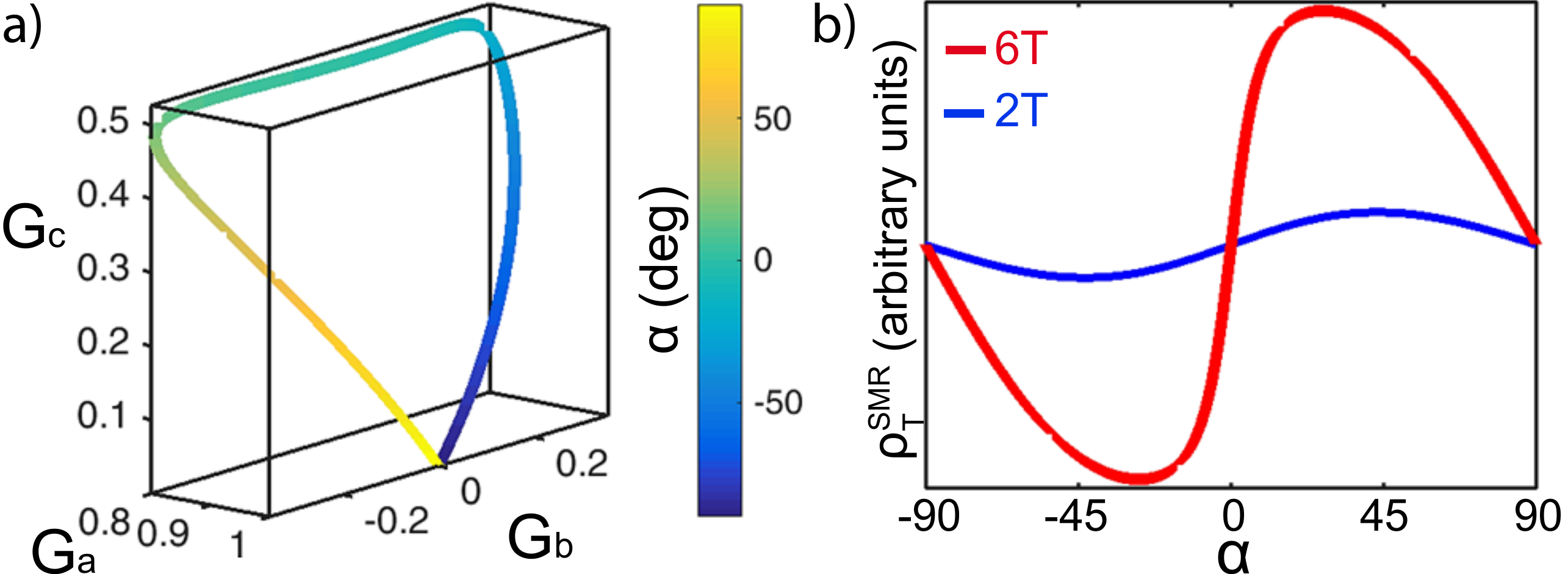} \caption{N\'{e}el vector,
$\mathbf{G}=\left(  G_{a},G_{b},G_{c}\right)  $ with $\left\vert
\mathbf{G}\right\vert =1$, calculated as a function of the magnetic field in
the $ab$ plane. The angle $\alpha\in\left[  -90{^{\circ},}90{^{\circ}}\right]
$ as defined in Fig.\thinspace\ref{fig:sample}(b) is coded by the colored bar.
$\mathbf{G}\left(  \alpha\right)  $ minimizes the free energy Eq.\thinspace
(\ref{eq:df}), for $K_{b}=0.15\,\mathrm{K}$ per Fe ion and $H=6\,\mathrm{T}$
(other parameters are given in the text). (b) The transverse SMR (arbitrary
units) due to the magnetic Fe sublattice for \thinspace$H=6\,\mathrm{T}$, i.e.
the $\mathbf{G}\left(  \alpha\right)  $ from panel (a) (thick red line), and
for $H=2\,\mathrm{T}$ (thin blue line).}%
\label{fig:rhoFe}%
\end{figure}

The orientation of the N\'{e}el vector $\mathbf{G}$ of the Fe sublattice at
temperatures well above $T_{\mathrm{N}}^{\mathrm{Dy}}$ is governed by several
competing interactions: (a) the magnetic anisotropy, which above the Morin
transition favors $\mathbf{G}\Vert\mathbf{a}$, (b) the Zeeman energy that
favors $\mathbf{G}\perp\mathbf{H}$ since the transverse magnetic
susceptibility of an AFM is higher than the longitudinal one, and
(c) the coupling of the WFM moment, $\mathbf{m}_{\mathrm{WFM}%
}\Vert\mathbf{a}$, to the applied magnetic field. This competition can be
described phenomenologically by the free energy density
\begin{equation}
f=\frac{K_{b}}{2}G_{b}^{2}+\frac{K_{c}}{2}G_{c}^{2}+\frac{\chi_{\perp}}%
{2}\left[  (\mathbf{G\cdot H})^{2}-\mathbf{H}^{2}\right]  -m_{\mathrm{WFM}%
}G_{c}H_{a}, \label{eq:df}%
\end{equation}
with the first two terms describing the second-order magnetic anisotropy with
magnetic easy, intermediate and hard axes along the $a$, $b$ and $c$ crystallographic directions,
respectively ($K_{c}>K_{b}>K_{a}=0$), $\chi_{\perp}$ is the transverse magnetic
susceptibility, and the $m_{\rm WFM}$ is the weak ferromagnetic moment along the $a$ axis, induced 
by $\mathbf{G}\Vert\mathbf{c}$. $\left\vert\mathbf{G}\right\vert =1$, 
because the longitudinal susceptibility of the Fe spins is very small for 
$T\ll T_{\mathrm{N}}^{\mathrm{Fe}}$. The magnetic
field $\mathbf{H}$ is chosen parallel to the $ab$ plane, but $\mathbf{G}$ can
have an OOP component $G_{c}\neq0$ since the third term in
Eq.(\ref{eq:df}) couples  $G_c$ linearly to $H_{a}$. For the SMR at 250~K, we may
disregard higher-order magnetic anisotropies that become important near
the Morin transition.

At weak magnetic fields, the magnetic anisotropy pins the N\'{e}el vector to the $a$
axis. When the Zeeman energy becomes comparable with the anisotropy energy,
the rotation of the magnetic field vector in the $ab$ plane gives rise to a
concomitant rotation of $\mathbf{G}$. In the absence of magnetic anisotropy,
the canting of the magnetic moments leads to $\mathbf{G}\perp\mathbf{H}$ 
for any magnetic field orientation due to the Zeeman energy rendering a sinusoidal SMR, 
but magnetic anisotropy can distort the angular dependence. 
This behavior is further complicated by the WFM: for strong magnetic fields 
along the $a$ axis, the N\'{e}el vector tilts away from the 
$ab$ plane towards the $c$ axis, since the $c$-component of
$\mathbf{G}$ induces a WFM moment parallel to the applied magnetic field
\cite{Prelorendjo1980, Eremenko1987}. By contrast, $G_{b}$ does not give rise
to a weak FM moment, so the N\'{e}el vector returns into the $ab$ plane when we
rotate the magnetic field away from the $a$ axis. The equilibrium N\'{e}el
vector minimizes the free energy Eq. (\ref{eq:df}) under the constraint
$\left\vert \mathbf{G}\right\vert =1$ as a function of strength and
orientation of the magnetic field with in-plane (IP) angle $\alpha$ (see Fig.
\ref{fig:sample}b)).

We adopt a weak magnetization parameters $m_{\mathrm{WFM}}=0.133\,\mu_{B}$ per
$\mathrm{Fe}^{3+}$ ion induced either by $\mathbf{G}\Vert\mathbf{c}$ along the
$a$ axis \cite{Cao2016} or by $\mathbf{G}\Vert\mathbf{a}$ along the $c$ axis
\cite{zvezdin1985}. The transverse magnetic susceptibility can be estimated using
the Heisenberg model with an Fe-Fe exchange constant $J_{1}=4.23\,$meV for
Y$_{3}$Fe$_{5}$O$_{12}$ \cite{Hahn2014a} 
, which leads to $\chi_{\perp}=\mu_{B}^{2}/(3J_{1})$, which does not depend strongly on the rare-earth ion.
$K_{c}$ governs the critical field when applied along the $a$ axis with
$\mu_0 H_{\mathrm{cr}}=9.3\,$T at $T=270$ K \cite{Prelorendjo1980}
that fully rotates $\mathbf{G}$ from the $a$ to the $c$ direction. 
$K_c$ can then be estimated using $K_{c}=m_{\mathrm{WFM}}H_{\mathrm{cr}}+\chi_{\perp
}H_{\mathrm{cr}}^{2}$
. $K_{b}$ is the only free temperature-dependent parameter that we fit to the 
field-dependent SMR. All other constants are taken to be independent of temperature. 
A typical calculated dependence of $\mathbf{G}\left(  \alpha\right)$ and the 
corresponding contribution of the Fe spins to the SMR is shown in Fig.~\ref{fig:rhoFe} 
(see below for a more detailed discussion).

\begin{figure}[t]
\includegraphics[width=8.4cm]{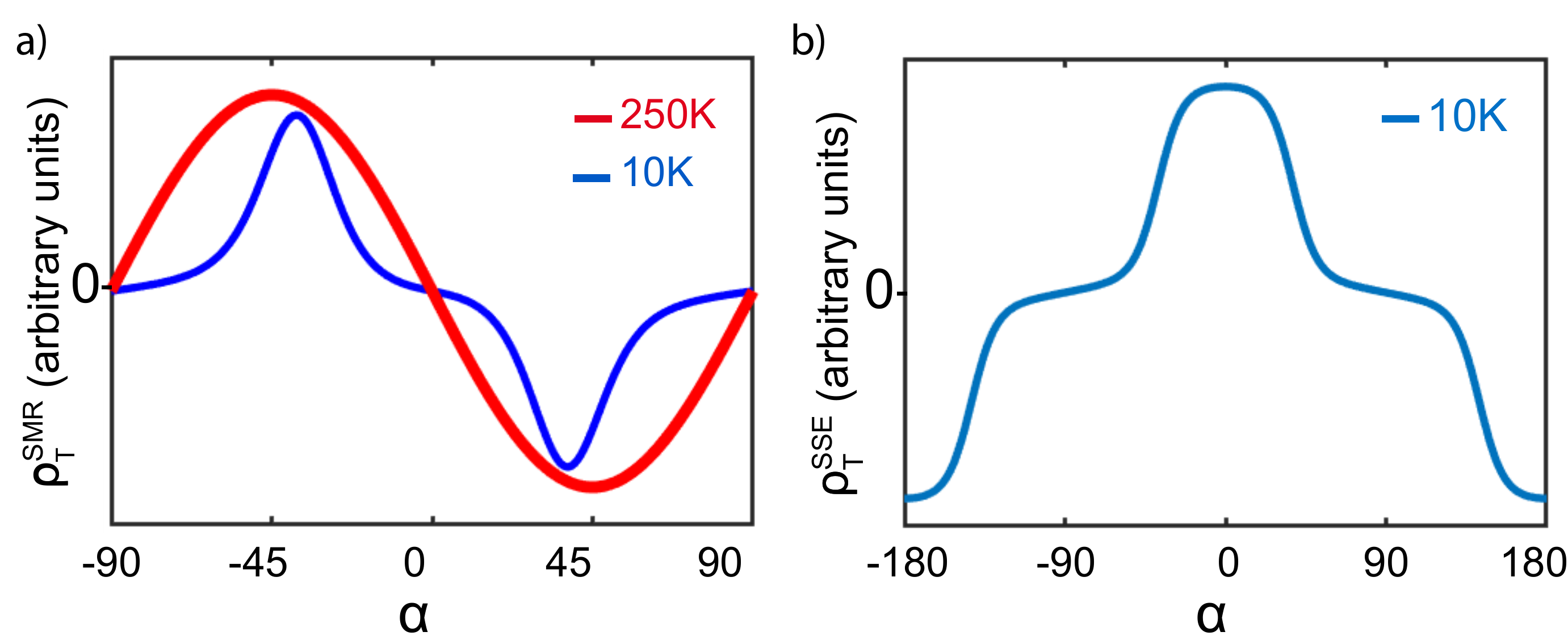}
\caption{Calculated angular dependence of the transverse a) SMR ($\rho
_{T}^{SMR}$) and b) local SSE ($\rho_{T}^{SSE}$) as contributed by
paramagnetic Dy$^{3+}$ moments polarized by an applied field $H=6\,$T. The
curve at 10$\,$K (blue line) is calculated numerically using Eq. \ref{eq:RT}.
The 250$\,$K curve (amplified by a factor 100, red line) is obtained
analytically from Eq. (\ref{HDY}). Both SMR and SSE grow with decreasing
temperature and associated increasing Dy$^{3+}$ magnetization. }%
\label{fig:rhoDy}%
\end{figure}

Ordered rare-earth ions can also contribute to the SMR and SSE. The spectrum
of the lowest-energy $^{6}$H$_{15/2}$ multiplet of the Dy$^{3+}$ ion (4f$^{9}$
electronic configuration) consists of a Kramers doublet separated by $\Delta=52$
cm$^{-1}\left(\approx 75\,\mathrm{K}\right)  $ from the
first excited state \cite{Zvezdin1979}. At low temperatures, $k_{\mathrm{B}%
}T\ll\Delta$, the Dy moments behave as Ising spins tilted by an angle $\pm
\phi_{\mathrm{Dy}}$ away from the $a$ axis in the $ab$ plane ($\phi_{\mathrm{Dy}%
}=57^{\circ}$). At high temperatures, $k_{\mathrm{B}}T\gg\Delta$, they can be
described as anisotropic Heisenberg spins with paramagnetic susceptibilities,
$\chi_{\parallel}^{\mathrm{Dy}}\left(  \chi_{\perp}^{\mathrm{Dy}}\right)  $
for a magnetic field parallel (perpendicular) to the local spin-quantization
axis ($\chi_{\parallel}^{\mathrm{Dy}}>\chi_{\perp}^{\mathrm{Dy}}$)
\cite{valiev2003}. 

For $k_{\rm B} T \gg \Delta$, the SMR resulting from the contributions of the four Dy sublattices (four Dy sites in the crystallographic unit cell of DFO) is
\begin{align}
R_{T}^{\mathrm{SMR}}  &  \propto-A\left[  H^{2}\sin(2\alpha)-2Hg_{1}G_{c}%
\sin\alpha\right] \nonumber\\
&  -2BHg_{2}G_{c}\sin\alpha, \label{eq:RT}%
\end{align}
where the first term originates from the interaction of Dy spins with the applied magnetic field and the other two terms result from the exchange field induced by Fe spins on Dy sites (for a more detailed discussion of the effective magnetic field acting on Dy spins and the expressions for A and B in terms of the magnetic susceptibilities of the Dy ions see Appendix \ref{Appendix B}).
It can be inferred form Fig. \ref{fig:rhoFe}\thinspace a) that
$G_{c}$ is approximately proportional to $\cos\alpha$. Therefore, all terms in Eq. \ref{eq:RT}
give the $\sin(2\alpha)$ dependence of the
transverse SMR at high temperatures (thick red line in Fig.~\ref{fig:rhoDy}~
a)). Equation \thinspace(\ref{eq:RT}) should be added to the SMR\ caused by the iron
sublattice with an unkown weight that is governed by the mixing conductance of
the Dy sublattice. We may conclude however that an additional $\sin(2\alpha)$
should not strongly change the shape of the SMR\ in Figure \ref{fig:rhoFe}b).

At low temperatures, , $T \ll \Delta / k_{\rm B}$, 
the Dy moments behave as Ising spins. A rotation of the magnetic field in the
$ab$ plane modulates the projection of the effective magnetic field on the local spin-quantization axes of the four Dy sublattices, which affects the angular dependence of  the SMR. Since the paramagnetic model Eq. \thinspace(\ref{eq:RT}) cannot be used anymore, we compute the Dy contribution to the SMR $\sim$ $m_{x}m_{y}$ numerically for the rare-earth Hamiltonian
\begin{equation}
H_{\mathrm{Dy}}^{(i)}=g_{J}\mu_{\mathrm{B}}(\mathbf{J}\cdot\mathbf{H}%
_{\mathrm{Dy}})-\frac{K}{2}(\mathbf{J}\cdot\hat{\mathbf{z}}_{i})^{2},\quad
i=1,2,3,4,\label{HDY}%
\end{equation}
with $\mathbf{J}$ as the Dy total angular momentum, $g_{J}=4/3$ the Land\'{e}
factor, $K=\Delta/7$ the anisotropy parameter, which is known to reasonably describe
the low-energy excited states of Dy ions and $\hat{{\bf z}}_i$ are the local easy 
axes rotated by $+ 57^\circ$, for the Dy sublattices 1 and 3, and  $- 57^\circ$, 
for the  sublattices 2 and 4, away from the $a$ axis. The magnetic field 
${\bf H}_{\rm Dy}$ acting on Dy spins is the sum of the 
applied field and the exchange field from Fe spins: ${\bf H}_{\rm ex} = 
g_1 G_z \hat{{\bf a}} \pm g_2 G_z \hat{{\bf b}}$, where the $+/-$ is for 
the sublattices $1,3$ and $2,4$, respectively. We neglect the $c$ component 
of the exchange field, since the Dy magnetic moment along the $c$ is small 
and does not affect the SMR. Using the Hamiltonian Eq. (\ref{HDY}), we 
calculate the average $a$ and $b$ components of the magnetic moments of 
the 4 Dy sublattices at a temperature $T$ and the resulting contributions to 
SMR. The angular dependence of the SMR due to Dy spins is plotted in 
Fig.~\ref{fig:rhoDy}~a). 

The calculations recover the $\sin(2\alpha)$ angular
dependence of the SMR from Eq.\thinspace(\ref{eq:RT}) at high temperatures.
At 10$\,$K (blue line) the SMR curve becomes strongly deformed: The angular 
dependence of the SMR shows peaks and dips at the effective field directions 
orthogonal to the quantization axis $\hat{\mathbf{z}}_{i}$ of the $i$-th 
rare-earth sublattice.

For long magnon relaxation time, the SSE generated a spin
current that is assumed to be proportional to the bulk magnetization and
can therefore provide additional information. We focus here on the low
temperature regime because we did not observe an SSE at elevated temperature,
which is an indication that the Dy magnetization plays an important role. \ 

A net magnetization of rare-earth moments affects the SSE signals
in gadolinium iron \cite{Geprags2016} and gadolinium gallium 
\cite{Oyanagi2019} garnets. We assume that the SSE is dominated by a spin
current from the bulk that is proportional to the total magnetization
$\mathbf{m}_{b}^{Dy}$ of the four Dy sublattices that we calculated for the
Hamiltonian Eq.\thinspace(\ref{HDY}) at 10\thinspace K as function of the 
angle $\alpha$ of the applied magnetic field. The model predicts peaks at
magnetic field directions aligned with the Ising-spin axes of the Dy moments, i.e. in between those canted by $\pm33\degree$, 
which enhances the magnetization. The
contribution from the Fe sublattice to the SSE is expected to depend as
$\cos{\alpha}$ on the external magnetic field direction \cite{Yuan2018}. The ratio of the Fe and Dy contributions to SSE is unknown.

\begin{figure}[t]
\includegraphics[width=8.4cm]{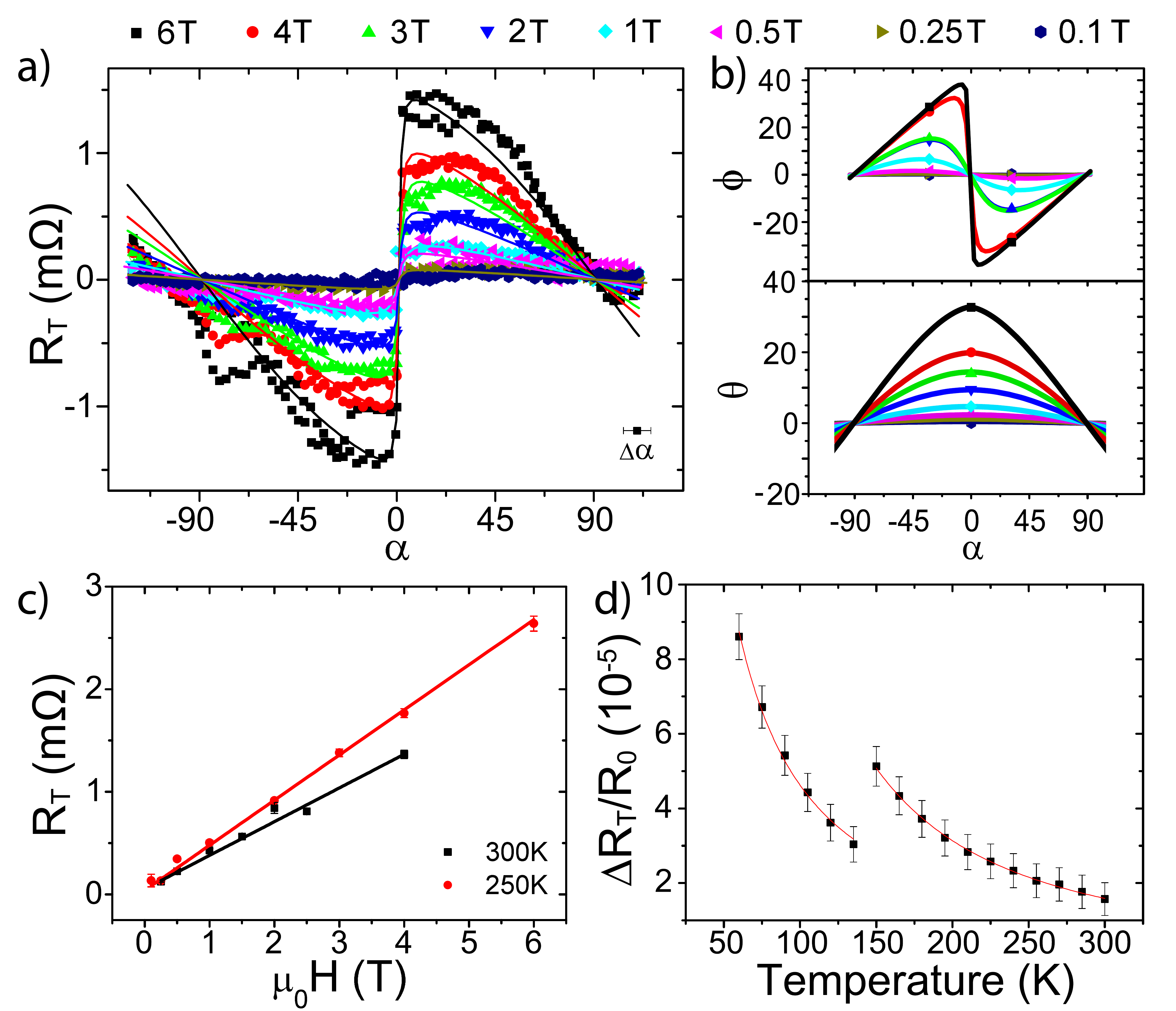}\caption{ (a)
Transverse SMR (symbols) measured as a function of IP magnetic field angle
$\alpha$ and strength (indicated at the top). The measurements are done on
device 1 with a current of 1$\,$mA at 250$\,$K and the error bar $\Delta
\alpha$ indicates a systematic error due to a possible misalignment of the
magnetic field direction as compared to the crystallographic axes. The lines
are fits obtained by adjusting $K_{b}$ in the free energy model Eq.
(\ref{eq:df}). (b) The IP ($\phi$) and OOP ($\theta$) canting angles of the
N\'{e}el vector with respect to $\mathbf{b}$ as a function of the IP
magnetic field direction from the fits. (c) The maximal signal change $\Delta R_{Tr}$  during a magnetic field rotation depends linearily on the magnetic field strength and (d) shows a power-law temperature dependence, $\Delta R_{Tr}/R_{0}\propto(T)^{\epsilon}$. $R_{0}$ is the sheet
resistance obtained from the base resistance of the corresponding longitudinal
measurements adjusted by the geometrical factor length/width of the Hall bar.
These measurements are carried out at 4$\,$T.}%
\label{fig:field-rot}%
\end{figure}

\begin{figure*}[t]
\includegraphics[width=17.8cm]{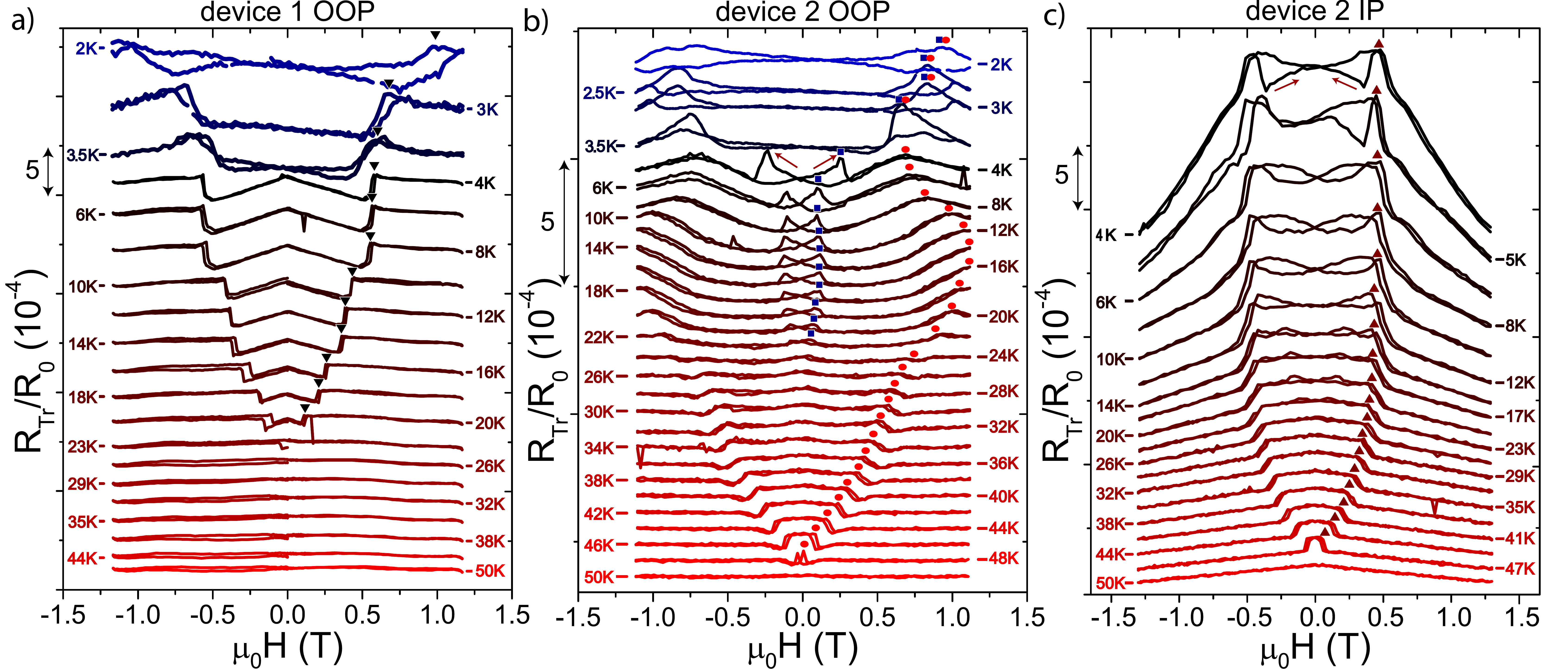}\caption{The relative changes in the transverse resistances 
$R_{\mathrm{Tr}}/R_{0}$ of (a) devices 1 and (b,c) device 2. 
A linear contribution from the ordinary Hall effect has been 
subtracted from the OOP data. Offsets of
the order of 10$^{-4}$ are removed and the curves are shifted with respect to
each other for clarity. The magnetic field directions 
are (a,b) along $\mathbf{z}$ for the OOP and (c) along $\mathbf{y}$ for the IP configurations. (a) The data for device 1 are expected to not change during
the Morin transition. The observed SMR is symmetric with respect to current
and magnetic field reversal and sensitive to Dy$^{3+}$ ordering. (b,c) Device 2 reveals the Morin transition by a positive step for weak magnetic
fields. Below 23$\,$K, hysteretic resistance features emerge when sweeping the
fields back and forth that vanishes at higher magnetic fields and
temperatures. The arrows indicate the magnetic field sweep directions, while
the symbols highlight the critical magnetic fields as summarized in Fig.
\ref{fig:DyFeO3_phase_boundaries}.}%
\label{fig:transverse_field-sweep}%
\end{figure*}

\section{Results}

\label{Results}

The SMR was measured by rotating an IP magnetic field of various
strengths. Temperature drift and noise swamped the small signal in the
longitudinal resistance as discussed in Appendix \ref{Appendix A}. Figure
\ref{fig:field-rot} a) shows the measured resistance of device 1 at 250$\,$K
in the transverse (planar Hall) configuration using the left contacts in Fig.
\ref{fig:sample} b). The results for the right Hall contacts (not shown) are
very similar.

The (negative) sign of the SMR agrees with our Fe sublattice model, suggesting
that it is caused by the AFM ordered Fe spins with N\'{e}el
vector $\mathbf{G}$ normal to the applied magnetic field. However,
$\mathbf{G}$ cannot be strictly normal to the magnetic field, because the SMR is not
proportional to $\sin{(2\alpha)}$, as observed for example in NiO
\cite{Hoogeboom2017}. The strongly non-sinusoidal angular dependence of the
SMR is evidence for a non-trivial path traced by the N\'{e}el vector in an
applied magnetic field as predicted by the model Eq.\thinspace(\ref{eq:df}).

\begin{figure}[t]
\includegraphics[width=8.3cm]{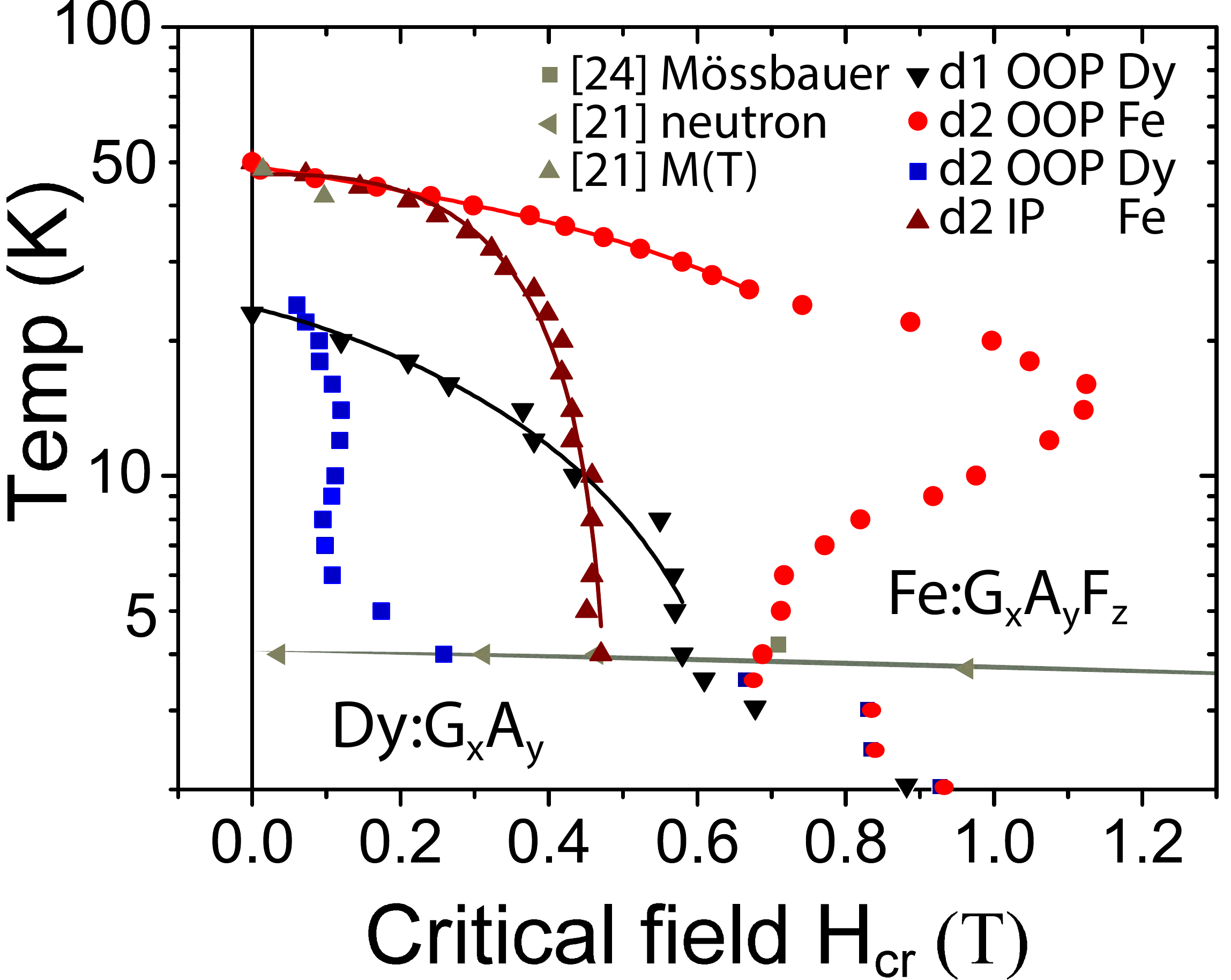}\caption{Critical
magnetic fields $H_{\mathrm{cr}}$ of the observed transitions in the transverse
resistance as a function of temperature. Symbols correspond to Fig.
\ref{fig:transverse_field-sweep}, where they denote the step functions that
trace the Morin transition in device 2.
$\textcolor{darkred}\filledmedtriangleup$ indicates IP and
{\protect\Large \textcolor{red}{$\bullet$}} OOP magnetic field directions. The
latter symbol describes the peaks at lower temperatures as well. The OOP
$H_{\mathrm{cr}}$ of the low magnetic field features are shown for device 1
($\filledmedtriangledown$) and device 2
(\textcolor{blue}{$\filledmedsquare$}). The features for the IP magnetic field
directions are less pronounced and not shown. The lines show a fit by the
function $H_{\mathrm{c}}\propto(T_{M}-T)^{\epsilon}$, which is used to extract
the ordering temperatures of 50$\,$K and 23$\,$K for the Morin transition and
a magnetic phase transition to an ordered Dy$^{3+}$ sublattice, respectively. 
Further data is from Refs. \cite{Wang2016,Prelorendjo1980}, obtained by M\"ossbauer spectrometry (\textcolor{greay}{$\filledmedsquare$}), neutron scattering (\textcolor{greay}{$\blacktriangleleft$}) and magnetometry (\textcolor{greay}{$\blacktriangle$}).
}%
\label{fig:DyFeO3_phase_boundaries}%
\end{figure}

Figure~\ref{fig:rhoFe} a) shows the dependence of the three components $G_{a}$,
$G_{b}$ and $G_{c}$ of the N\'{e}el vector on the IP orientation angle
$\alpha$ of the magnetic field, for $\mu_0 H=6$ T. The value of $\alpha\in\left[
-90^{\circ},90^{\circ}\right]  $ is indicated by the color code side bar. When
$\alpha=0$ $(\mathbf{H}\Vert\mathbf{a})$, the magnetic field causes a tilt of
$\mathbf{G}$ away from the easy $a$ axis towards the hard $c$ axis since the
N\'{e}el vector parallel to the $c$ axis induces a magnetization along the $a$
axis. The excursion of $\mathbf{G}$ from the $ab$ plane effectively reduces
the role of the IP magnetic anisotropy, which leads to a large rotation
of the N\'{e}el vector in the $ab$ plane for small $\alpha$ (at nearly
constant $G_{c}$). As explained above, this rotation is driven by the Zeeman
energy of the AFM ordered Fe spins (the third term in
Eq.(\ref{eq:df})), which favors $\mathbf{G}\perp\mathbf{H}$ and competes with
the magnetic anisotropy that favors $\mathbf{G}\Vert\mathbf{a}$ (the first
term in Eq.(\ref{eq:df})). This behavior is similar to the spin-flop
transition for a magnetic field applied along the magnetic easy axis, except that
$\mathbf{G}$ does not become fully orthogonal to the magnetic field. As the
magnetic field vector rotates away from the $a$ axis, $G_{c}$ and $\left\vert
G_{b}\right\vert $ decrease, and at $\alpha=\pm90^{\circ}$, $\mathbf{G}$ is
parallel to the $a$ axis.

The sensitivity of $\mathbf{G}$ to small $\alpha$ gives rise to an abrupt
change of the transverse SMR that is proportional to $G_{a}G_{b}$ close to
$\alpha=0$ (thick red line Fig.~\ref{fig:rhoFe}b). The calculated and observed
SMR scans agree well for $T=250\,$K and $\mu_0 H=6\,$T. Surprisingly, the shape of the
experimental curves is practically the same at all magnetic field strengths, i.e. the SMR
jumps at $\alpha=0$ even at weak fields, while the calculation approach the
geometrical $\sin(2\alpha)$ dependence (thin blue line in Fig.~\ref{fig:rhoFe}%
b) calculated for $\mu_0 H=2$ T). The fits of the observed SMR for all magnetic fields
require a strongly field-dependent IP anisotropy parameter $K_{b}$ that
is very small in the zero field limit: $K_{b}=\left(  6\pm8\right)
\cdot10^{-6}+\left(  3.20\pm0.02\right)  \cdot10^{-3}$\thinspace$\left(
H/\mathrm{T}\right)  \ \mathrm{^{2}}$ K (see Fig.~\ref{fig:field-rot}a). At
present we cannot explain this behavior. The Dy$^{3+}$ moments should not play
an important role in this regime unless a Pt induced anisotropy at the DFO/Pt
interface modifies their magnetism (see below).

The exchange coupling between the rare-earth and transition-metal magnetic
subsystems is reflected by the second term in Eq.(\ref{eq:RT}) of the
Dy$^{3+}$ contribution to the SMR that is proportional to $G_{c}$, i.e. the
AFM order of the Fe spins. Since, $G_{c}$ is a smooth function
at $\alpha=0$, it cannot be hold responsible for the large zero-field
magnetoresistance. The angular SMR appears to be dominated by the N\'{e}el
vector $\mathbf{G}$ of the Fe moments, in contrast to SmFeO$_{3}$, in which
the Sm-ions determine not only the amplitude but also the sign of the SMR
\cite{Hajiri2019}.

The linear increase of the SMR with magnetic field strength (see Fig.
\ref{fig:field-rot}c)) can partly be explained by the growth of the maximum
IP rotation angle, $\phi$, of the N\'{e}el vector with magnetic field. However,
deviations from the linear dependence are then expected close to the critical
value, $H_{a}\sim9$ T, at which the re-orientation transition from
$\mathbf{G}\Vert\mathbf{a}$ to $\mathbf{G}\Vert\mathbf{c}$ in $\mathbf{H}%
\Vert\mathbf{a}$ is complete \cite{Prelorendjo1980}. Nevertheless, the SMR
signal shows no sign of saturation at $\mu_0 H_{a}=6\,$T and $T=250\,$K. The $\mu_0H$ of Dy becomes of the order of k$ _B$T at a magnetic field strength of 37$\,$T, indicating contributions from the paramagnetic rare earth spins remains linear in the applied field strengths.


Further evidence for rare earth contributions at higher temperatures is the
Curie-like power-law temperature dependence of the SMR (see Fig.
\ref{fig:field-rot}d)) $\mathrm{SMR}\sim T^{\epsilon}$, with $\epsilon
=-1.24\pm0.04$ at low temperatures and $\epsilon=-1.67\pm0.02$ at high
temperatures.\footnote{We have not been able to identify the mechanism for the
step observed between 135$\,$K and 150$\,$K that has to our knowledge not been
reported elsewhere either.} For comparison, in the AFM NiO,
$\epsilon$ is positive and the SMR signal grows quadratically with the
AFM order parameter \cite{Hoogeboom2017}. At temperatures well
below the N\'{e}el transition $T_{N}^{\mathrm{Fe}}=645$ K, the Fe based
magnetic order is nearly temperature independent. The strong magnetic field
and temperature dependence therefore suggest important contributions from
polarized Dy$^{3+}$ moments even at room temperature.

The puzzling strong magnetic field-dependence of $K_{b}$ from the data fit  might indicate a different coupling between the rare earth and
transition metal magnetic subsystems at the interface and in the bulk. It can
be justified by the following symmetry argument.\textit{ }The generators of
the Pbnm space group of the DFO crystal are three (glide) mirror
planes: $\tilde{m}_{a}$, $\tilde{m}_{b}$ and $m_{c}$, i.e. a mirror reflection combined with a shift along a direction parallel to the mirror plane.  $m_{c}$ is
broken at the interface normal to the $c$ axis. In the absence of $m_{c}$, the
rare earth order parameters $A_{a}^{\prime}$ and $G_{b}^{\prime}$ transform to $G_{b}$ that describes the AFM order of Fe spins, which allows for a
linear coupling between the rare earth and Fe spins at the interface. Since $G_{b}$ strongly depends on $\alpha$
at $\alpha=0$, the same may hold for the rare earth moments at the interface.
The SMR is very surface sensitive and could be strongly affected by this
coupling.

Next, we turn to the SMR at temperatures below the Morin transition at
magnetic fields around the re-entrant field, $H_{\mathrm{cr}}$.
Figure~\ref{fig:transverse_field-sweep}(a) shows the transverse SMR of device
1 in an OOP magnetic field, while the data for longitudinal resistance are
deferred to the Appendix \ref{Appendix A}, Fig.~\ref{fig:DyFeO3_long}a). 
We subtracted a linear field dependent contribution from the OOP data 
that is caused by the ordinary Hall effect in Pt.

The zero-field resistance of device 1 should not change under the Morin
transition when the N\'{e}el vector direction switches from $a$ to $b$ nor
should it be affected by weak magnetic fields $\mathbf{H}\Vert\mathbf{c}$
($\mu_{0}H_{\mathrm{cr}}<0.1\,$T near 50$\,$K \cite{Wang2016}) that return the
system to $\mathbf{G}\Vert\mathbf{a}$. Indeed, we do not see any weak-field
anomaly of the SMR near 50 K in Fig.~\ref{fig:transverse_field-sweep}a).
However, below 23$\,$K, a negative SMR proportional to the applied field
appears. The linear field-dependence ends abruptly with a positive step-like
discontinuity (see Fig.~\ref{fig:transverse_field-sweep}a)). No resistance
offset has been observed between the zero-field $\Gamma_{1}$ and the
high-field $\Gamma_{4}$ phases. After substraction of the strictly linear
ordinary Hall effect contribution, the SMR feature is an even function of
$H_{\mathrm{c}}$. The magnetic phase transition at 23$\,$K appears to be
unrelated to the Morin transition and has not been reported previously.

The Morin transition is clearly observed in the OOP and IP SMR of device 2, in
which the crystallographic axes are azimuthally rotated by $45{{}^{\circ}}$
relative to the Hall bar as shown in Fig. \ref{fig:sample}b). Here, an SMR
signal is expected for both magnetic phases and the $90{{}^{\circ}}$
rotation of the N\'{e}el vector from $a$ to $b$ should change its sign from
positive for the AFM $\Gamma_{1}$ phase ($\mathbf{G}\parallel\mathbf{b}$) to
negative for the WFM $\Gamma_{4}$ phase ($\mathbf{G}\parallel\mathbf{a}$), for
$\left\vert H_{c}\right\vert >H_{\mathrm{cr}}$. The $\Gamma_{1}$ phase can
also be suppressed by an IP field $\mathbf{H}\parallel\hat{y}=\hat{b}-\hat{a}$
that rotates the N\'{e}el vector towards $\hat{b}$ to lower the Zeeman energy.
The drop in the Hall resistance observed in device 2 below 48$\,$K for the OOP
(Fig.~\ref{fig:transverse_field-sweep} b)) and IP
(Fig.~\ref{fig:transverse_field-sweep} c)) field directions can therefore be
ascribed to the Morin transition with a temperature-dependent $H_{\mathrm{cr}%
}$. The SMR steps are negative, as expected.

At even lower temperatures the model appears to break down since we observe
hysteretic behavior in the field-dependence of the SMR signal at low magnetic
fields for both the OOP and IP directions. These features come up below
23$\,$K, so appear to have the same origin as the anomalies in device 1. For
the OOP direction, the low-field anomalies in device 2 are peaks while they
are step-like in device 1. Wang et al.\cite{Wang2016} did not observed a
hysteresis in the Fe$^{3+}$ magnetic sublattice and suggested that observed
hysteretic behaviour \cite{Tokunaga2008,Li2014} is an evidence for long-range to
short-range Dy$^{3+}$ magnetic order. The SMR might witness an ordering of
Dy$^{3+}$ moments at the interface at a higher temperature than in the bulk
that cannot be detected by other measurements.

Another unexpected feature is a linear negative magnetoresistance at
$\left\vert H_{y}\right\vert >H_{\mathrm{cr}}$ for the IP configuration (see
Fig. \ref{fig:transverse_field-sweep}c)) that might be caused by a canting of
$\mathbf{G}^{\mathrm{Fe}}$ towards $\mathbf{c}$ by $H_{\mathrm{a}}>1.6\,T$
\cite{Prelorendjo1980}. A misalignment of the crystallographic axes could also
affect the SMR more significantly for high magnetic fields. However, neither
of these mechanisms explain the IP magnetic field dependence and the peaks and
low magnetic field features in the OOP measurements of both devices below
23$\,$K (Fig. \ref{fig:transverse_field-sweep}a) and
\ref{fig:transverse_field-sweep}b)). Since their signs and shapes vary, we can
exclude a paramagnetic OOP canting of the Dy$^{3+}$ orbital moments. The
Dy$^{3+}$ orbital moments are locked to the Ising axis in the $ab$ plane and
the magnetization is one order of magnitude larger in this plane than along
the $c$ direction \cite{Tokunaga2008}. This might explain the IP SMR features
in terms of an IP field and temperature dependent order of the Dy$^{3+}$
moments.

\begin{figure}[ptb]
\caption{The SSE, i.e. the detected voltage in the transverse Hall probe divided by the squared current of device 1 at 10$\,$K as a function of the magnetic field
strength and direction $\alpha$. At weak fields, the SSE shows a $\cos\alpha$
dependence as expected for the Fe$^{3+}$ magnetic sublattice. This amplitude
initially increases with the magnetic field strength but decreases again and
flattens for $H>0.5\,$T.}%
\includegraphics[width=7.0cm]{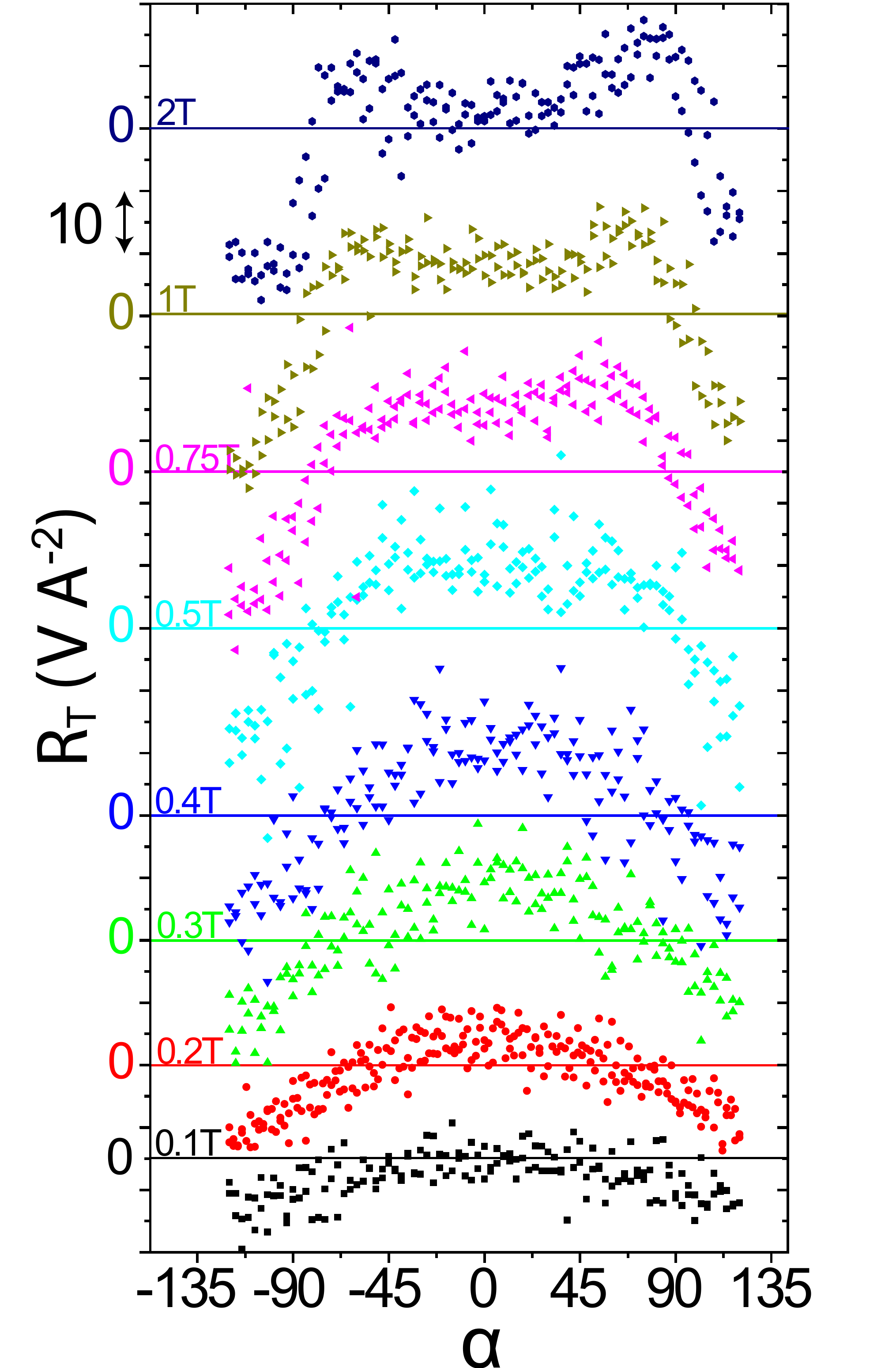}\label{fig:SSE}%
\end{figure}

The 90${^{\circ}}$ spin reorientation at the Morin transition maximizes
the Fe$^{3+}$ contribution to the SMR. The increase of the IP signal amplitude by
one order of magnitude upon lowering the temperature, see Fig.
\ref{fig:transverse_field-sweep}(c) is therefore unexpected. The signals
become as large as 1\%, one order of magnitude larger than the SMR signals of Pt on Y$_3$Fe$_5$O$_{12}$ \cite{Nakayama2013, Bauer2013, Vlietstra2013, Althammer2013} and a factor four larger than that of $\alpha$-Fe$_{2}$O$_{3}$ \cite{Fischer2019}. Ordered Dy$^{3+}$ magnetic moments appear to be responsible for the anomalous signals below 23$\,$K. They interact with the Fe sublattice by the exchange interaction, as observed before in the multiferroic phase at temperatures
exceeding $T_{N}^{\mathrm{Dy}}$ under a 0.5$\,$T magnetic field
\cite{Wang2016}. A contribution of Dy$^{3+}$ moments to the magnetization has
also been observed in terms of an upturn of the magnetization and hyperfine
field below 23$\,$K \cite{Reddy2015}.

The SMR steps in device 1 around $T_{N}^{\mathrm{Dy}}=4\,$K at which the Dy
moments order spontaneoulsy, are similar to those at higher temperature, which
supports the hypothesis that the latter are also related to Dy$^{3+}$
order.\textit{ }Device 2 shows an increased H$_{\mathrm{cr}}$ matching those
in device 1 at these temperatures. Both devices show no non-linear
antisymmetric field dependence, indicating that the Dy$^{3+}$ ordering above
4$\,$K is field-induced. Li et al. \cite{Li2014} observed jumps in the thermal
conductivity around 4$\,$T and attributed these to a spin reorientation of the
Fe sublattice. However, no further transitions are observed up to 6$\,$T as is
shown 
in Appendix \ref{Appendix A}, so we cannot confirm such an Fe$^{3+}$ transition.

The magnetic field and temperature of the occurrences of SMR steps at spin transitions and of SMR anomalies are collected in Fig. \ref{fig:DyFeO3_phase_boundaries}, including the peaks in
the OOP measurements of device 2, using the same markers as in Fig.
\ref{fig:transverse_field-sweep}. The data on the Morin transition agrees with
previous observations \cite{Wang2016, Prelorendjo1980}. The Morin point for
both IP an OOP configurations is around 50$\,$K, whereas the transitions ascribed to
an ordering of the Dy$^{3+}$ moments occur around 23$\,$K. Upon lowering the
temperature, the transitions associated to the Dy$^{3+}$ and Fe$^{3+}$ moments
approach each other and merge below $T_{N}^{\mathrm{Dy}}$, which is another
indication of a strong inter-sublattice exchange interaction.

Figure \ref{fig:SSE} summarizes the observed IP SSE data of device 1 at 10$\,$K.
The angular dependence of the resistance at small fields shows the $\cos{\alpha
}$ dependence, indicating that the magnon spin current $\mathbf{j}_{m}$
injected into Pt is constant with angle. The amplitude initially increases linearly with
field, but decreases again for $H>0.5\,$T. The SSE signal of a uniaxial AFM has 
$\cos\alpha$ dependence for an IP rotating magnetic field \cite{Yuan2018}. 
The SSE is small at angles for which our model for the Dy$^{3+}$ contribution 
in Fig. \ref{fig:rhoDy}b) predicts a peak. However, we do not observe the expected 
Dy$^{3+}$-induced SSE contribution due to the Dy$^{3+}$ magnetization shown in Fig.
\ref{fig:rhoDy}. On the contrary, an increase in Dy$^{3+}$ magnetization appears 
to suppress the SSE signal. These results suggest that the angular dependence of the 
SSE is governed not so much by the ordering of the Dy spins, but by their effect 
on the frequencies of the antiferromagnons in the Fe magnetic subsystem. 
The ordering of Dy spins leads to a hardening of the AFM resonance modes 
\cite{Stanislavchuk2016}. The applied magnetic field suppresses the Dy spin 
ordering and results in a substantial decrease of the spin gap \cite{Stanislavchuk2016}
, which affects the thermal magnon flux and, hence, the SSE. At room temperature, 
the SSE signal does not rise above the noise level of 0.18$\,$V A$^{-2}$.


\section{Conclusion}

We studied the rare earth ferrite DFO by measuring the transverse
electric resistance in Pt film contacts as function of temperature and applied
magnetic field strength and direction. Results are interpreted in terms of 
SMR and SSE for magnetic configurations that minimize a magnetic free energy model with magnetic anisotropies, Zeeman energy and DMI. The N\'{e}el vector
appears to slowly rotate OOP and displays jumps under IP
rotating magnetic fields. Magnetic field-strength dependences indicate that
Fe$^{3+}$ spins are responsible for the symmetry of the SMR, but that the
Dy$^{3+}$ orbital moments affect the amplitude. The first-order Morin
transition is clearly observed at temperatures below 50$\,$K. Additional sharp
features emerge below 23$\,$K at critical fields below that of the Morin
transition. These observed features cannot be understood by the Fe$^{3+}$
N\'{e}el vector driven SMR. Rather, they suggest a magnetic field-induced
ordering of Dy$^{3+}$ established by the competition between applied magnetic
and exchange fields with Fe$^{3+}$. This hypothesis is supported by the
similar SMR features at the spontaneous Dy$^{3+}$ moment ordering temperature
$T_{N}^{\mathrm{Dy}}$. A Dy$^{3+}$ order above $T_{N}^{\mathrm{Dy}}$ also
appears to suppress the SSE contributions from the Fe sublattice.

Concluding, we report simultaneous manipulation and monitoring of the ordering
of both transition metal and rare earth magnetic sublattices and their
interactions as a function of temperature and magnetic field in the complex
magnetic material DFO.\newline

\section{Acknowledgements}

We thank A. Wu for growing the single crystal DyFeO$_{3}$, J. G. Holstein, H.
Adema, T. J. Schouten, H. H. de Vries and H. M. de Roosz for their technical
assistance as well as R. Mikhaylovskiy and A. K. Zvezdin for discussions. This
work is part of the research program Magnon Spintronics (MSP) No. 159 financed
by the Nederlandse Organisatie voor Wetenschappelijk Onderzoek (NWO) and JSPS
KAKENHI Grant Nos. 19H006450, and the DFG Priority Programme 1538 Spin-Caloric
Transport (KU 3271/1-1). Further, the Spinoza Prize awarded in 2016 to B. J. van Wees by NWO is gratefully acknowledged

\bibliographystyle{apsrev4-1}
\bibliography{references}

\begin{figure*}[t]
\includegraphics[width=17.8cm]{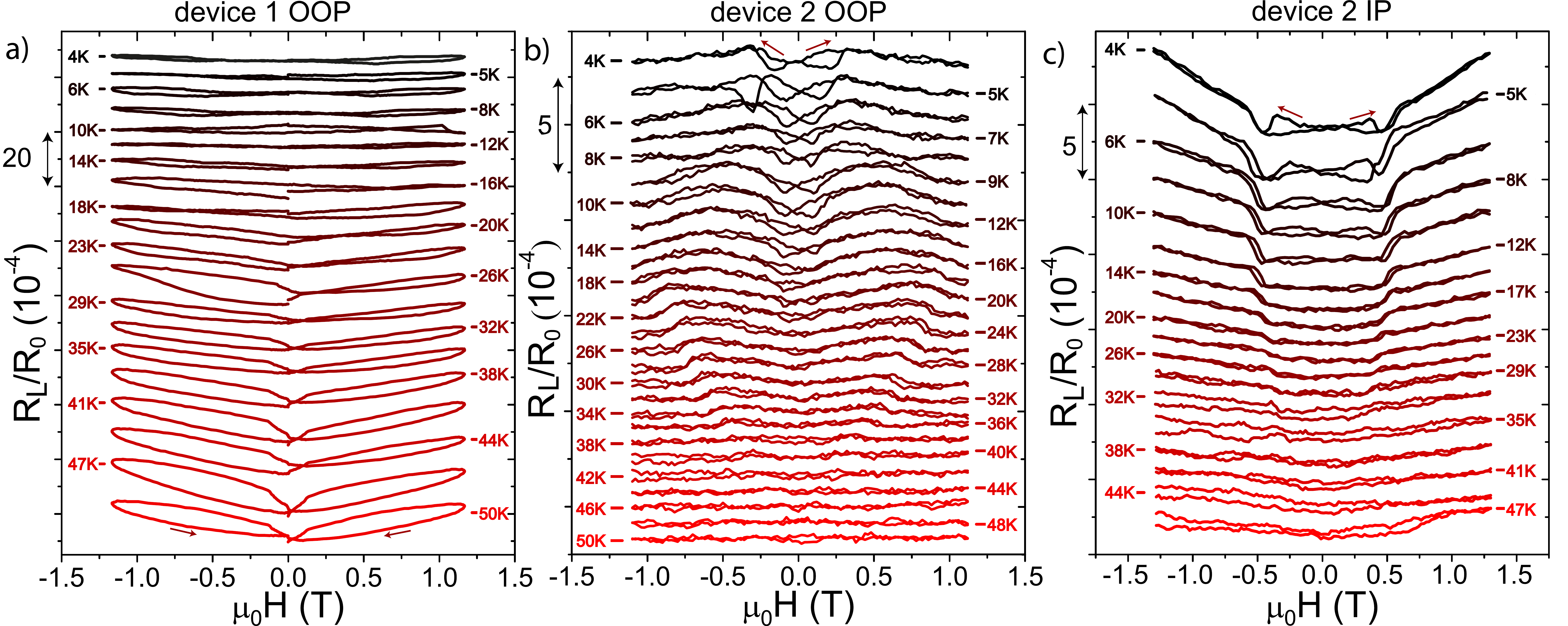}\caption{Relative changes in the longitudinal SMR $R_{L}/R_{0}$ 
of (a) device 1 as well as (b,c) device 2 at temperatures up to 50K for 
magnetic field sweeps (a,b) OOP and (c) IP.
Device 1 shows large hysteretic effects at the full range of magnetic field
strengths. The amount of data points around zero magnetic field, and thus the
waiting time per magnetic field change, is higher than at higher fields as to
have higher resolution for the transverse Morin transition. This makes the
hysteretic effects slightly distorted compared to a situation with constant
waiting time. Device 2 shows hysteretic effects solely at lower field
strengths which corresponds to the hysteretic features of the transverse
measurements.}%
\label{fig:DyFeO3_long}%
\end{figure*}

\begin{figure}[h]
\includegraphics[width=8.6cm]{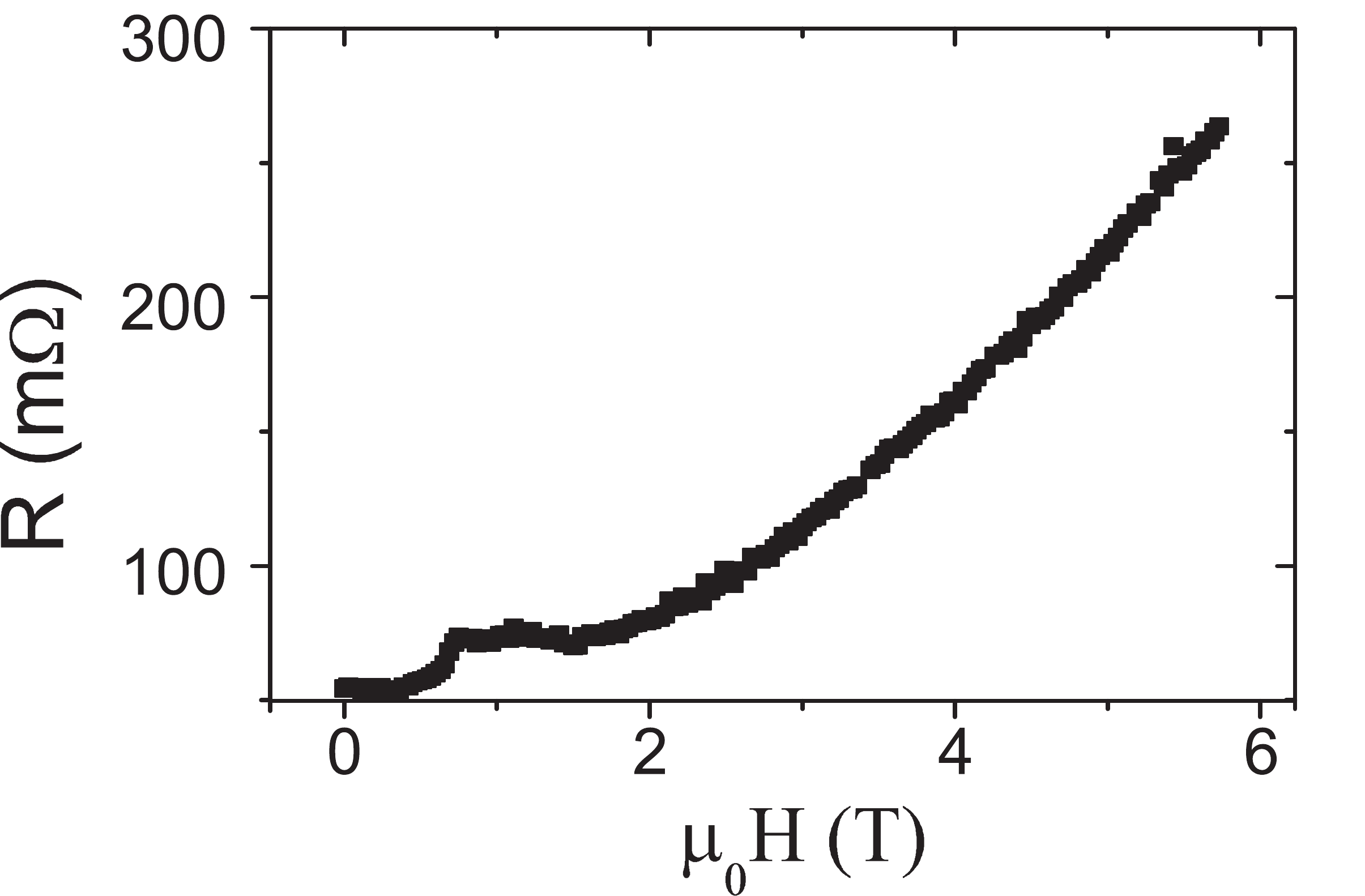} \caption{Transverse resistance
of device 2 at 2$\,$K as a function of the OOP magnetic field up to 6$\,$T.
The resistance increases continuously with magnetic field strength above
2$\,$T.}%
\label{fig:DyFeO3_temp}%
\end{figure}

\appendix

\section{Longitudinal and 2$\,$K SMR}

\label{Appendix A}

The modulation of the longitudinal Pt resistance as a function of magnetic
field are shown in Fig. \ref{fig:DyFeO3_long} for comparison with the
transverse SMR. The longitudinal signals are affected by a background contact
resistance that is sensitive to temperature changes. The SMR signals are
therefore more distorted by a small temperature drift than the transverse
measurements. Moreover, the background resistance suffer from increased noise.

The OOP resistance changes of device 1 are one order of magnitude larger than
those of device 2 and dominated by hysteretic effects. The signal amplitudes
of OOP and IP configurations for device 2 are similar. The measurement time of
one data point below 0.2$\,$T is smaller than at larger fields, influencing
the shape of the graphs. Device 2 shows hysteretic features at low magnetic
fields and below 23$\,$K, for both IP and OOP magnetic fields that are similar
to the transverse SMR features discussed in the main text.

Results of a field sweep up to 6$\,$T are shown in Fig. \ref{fig:DyFeO3_temp}.
The resulting continuous curve does not show transitions on top of those
discussed in the text, without evidence for a phase transition at 4$\,$T and
2$\,$K \cite{Li2014, Gnatchenko1994}.

\section{Exchange interaction}

\label{Appendix B}

The \textit{Pbnm} crystal symmetry allows an exchange coupling between
the Dy$^{3+}$ moments and G-type AFM ordered Fe spins. The coupling of 
the 4 (individual) Dy spins in the unit cell with the Fe spins is described as
\begin{align}
E_{\mathrm{Dy-Fe}}  &  =-g_{1}G_{c}\left( m_{1}^{a}+m_{2%
}^{a}+m_{3}^{a}+m_{4}^{a}\right) \nonumber\\
&  -g_{2}G_{c}\left( m_{1}^{a}-m_{2}^{a}+m_{3}^{a}-m_{4}^{a}\right)
\nonumber\\
&  -g_{3}G_{a}\left( m_{1}^{c}+m_{2}^{c}+m_{3}^{c}+m_{4}^{c}\right)
\nonumber\\
&  -g_{4}G_{b}\left( m_{1}^{c}-m_{2}^{c}+m_{3}^{c}-m_{4}^{c}\right)  ,
\label{eq:interaction}%
\end{align}
where the indices $1,2,3,4$ label the rare-earth ions in the unit cell. The exchange field from Fe ions is estimated to be $\sim2\,$T at low temperatures \cite{Zvezdin1979}.

For $k_{\mathrm{B}}T\gg\Delta$, the magnetization of the Dy sublattice
$m_{\parallel}=\chi_{\parallel}^{\mathrm{Dy}}H_{\parallel}$ and $m_{\perp
}=\chi_{\perp}^{\mathrm{Dy}}H_{\perp}$ for field components parallel and
perpendicular to the local anisotropy axis and $H=\sqrt{H_{\parallel}%
^{2}+H_{\perp}^{2}}$. We assume that the transverse SMR caused by the
paramagnetic Dy$^{3+}$ moments polarized by the applied field is proportional
to $m_{x}m_{y}$ \cite{Jiang2018, Han2014, Aqeel2015, Oyanagi2020}. Adding the
contributions of the four Dy sites in the crystallographic unit cell of
DFO and the exchange field from the Fe spins acting on the Dy spins as described in the main text, we obtain Eq. \ref{eq:RT} with $A=\left[  (\chi_{\parallel}^{\mathrm{Dy}}+\chi_{\perp}^{\mathrm{Dy}%
})^{2}-(\chi_{\parallel}^{\mathrm{Dy}}-\chi_{\perp}^{\mathrm{Dy}})^{2}%
\cos(4\phi_{\mathrm{Dy}})\right]  /2$ and $B=\sin(2\phi_{\mathrm{Dy}})\left[
(\chi_{\parallel}^{\mathrm{Dy}})^{2}-(\chi_{\parallel}^{\mathrm{Dy}}%
)^{2}-(\chi_{\parallel}^{\mathrm{Dy}}-\chi_{\parallel}^{\mathrm{Dy}}%
)^{2}\,\cos(2\phi_{\mathrm{Dy}})\right]  $. The coupling constants g$_3$ and g$_4$ do not appear in the expression for SMR since the latter does not depend on the c-component of Dy spins. Moreover, the c-component is very small at low temperatures, since the easy axes of Dy ions lie in the ab plane. Both g$_1$ and g$_2$ lead to (nearly) the same angular dependence of SMR. 




\end{document}